\documentclass[preprint,authoryear]{elsarticle}
\usepackage{amssymb,amsmath,graphicx,color}

\newcommand{\ave}[1]{\left\langle #1 \right\rangle}
\newcommand{\e}{\mathrm{e}}
\newcommand{\muc}{\mu_\mathrm{c}}
\newcommand{\muct}{\tilde\mu_\mathrm{c}}

\journal{Journal of Theoretical Biology}

\begin{document}

\begin{frontmatter}

\title{Error thresholds for self- and cross-specific enzymatic replication}
\author[lmu]{Benedikt Obermayer}
\ead{obermayer@physik.lmu.de}

\author[lmu]{Erwin Frey\corref{cor1}}
\ead{frey@physik.lmu.de}

\address[lmu]{Arnold-Sommerfeld-Center for Theoretical Physics and Center for NanoScience, Ludwig-Maximilians-Universit\"at M\"unchen, Theresienstr. 37, 80333 M\"unchen, Germany}
\cortext[cor1]{Corresponding author}

\begin{abstract}
The information content of a non-enzymatic self-replicator is limited by Eigen's error threshold. Presumably, enzymatic replication can maintain higher complexity,  but in a competitive environment such a replicator is faced with two problems related to its twofold role as enzyme and substrate: as enzyme, it should replicate itself rather than wastefully copy non-functional substrates, and as substrate it should preferably be replicated by superior enzymes instead of less-efficient mutants. Because specific recognition can enforce these propensities, we thoroughly analyze an idealized quasispecies model for enzymatic replication, with replication rates that are either a decreasing (self-specific) or increasing (cross-specific) function of the Hamming distance between the recognition or ``tag'' sequences of enzyme and substrate. We find that very weak self-specificity suffices to localize a population about a master sequence and thus to preserve its information, while simultaneous localization about complementary sequences in the cross-specific case is more challenging. A surprising result is that stronger specificity constraints allow longer recognition sequences, because the populations are better localized. Extrapolating from experimental data, we obtain rough quantitative estimates for the maximal length of the recognition or tag sequence that can be used to reliably discriminate appropriate and infeasible enzymes and substrates, respectively.
\end{abstract}


\end{frontmatter}

\section{\label{sec:intro}Introduction}

The acclaimed experimental finding~\citep{CechBR:90} that RNA not only stores genetic information but also provides catalytic function has inspired the RNA world scenario~\citep{GilbertNature:86}, a hypothesis for the starting point of Darwinian evolution at the origin of life through self-replication of RNA-like polynucleotides. Substantial progress over the last decades demonstrates the capability of RNA enzymes (ribozymes) to catalyze diverse chemical reactions~\citep{DoudnaNature:02,LilleyCOSB:05,JoyceAC:07}, among them the polymerization of as many as 20 nucleotides to a template molecule~\citep{JohnstonScience:01, ZaherRNA:07}, and even replication through template-directed ligation reactions involving short RNA oligomers as building blocks~\citep{LincolnScience:09}. However, the currently known systems are not yet capable of Darwinian evolution, lacking either the ability to replicate molecules as long and complex as themselves or to introduce heritable variation. 

For theorists, the focus has mainly been on whether the information content of a self-replicating molecule can be maintained in the presence of replication errors, usually employing Eigen's well-known quasispecies theory~\citep{EigenACP:89} for the self-replication of $L$-nucleotide sequences. In this model, replication errors occur with an error probability $\mu$ per single nucleotide, and the replication rates are taken as functions only of the template sequence. This mapping from genotype to replication rate (or fitness) presents a considerable challenge for evolutionary models, and even generic structural features of such fitness landscapes, such as the extent of ruggedness and epistasis, are under debate. Valuable insight comes from experiments on mutagenized ribozymes~\citep{KunNG:05} and from computer simulations using RNA secondary structure as phenotype~\citep{HuynenPNAS:96,TakeuchiBEB:05}, which indicate a significant degree of neutrality around an optimal sequence, reducing the deleterious effects of mutations. Other theoretical studies used a large number of different idealized fitness landscapes, e.g., with a single peak at one fittest ``master'' sequence~\citep{SwetinaBC:82,LeuthausserJCP:86,SchusterBMB:88,WoodcockJTB:96,GalluccioPRE:97,HermissonTPB:02,PelitiEPL:02,SaakianPNAS:06,SaakianPRE:09}, with some rather generic results~\citep{WieheGR:97,JainarXiv:05}: the population in sequence space is characterized by a broad mutant distribution (a quasispecies) localized about the master sequence for mutation probabilities smaller than a critical value $\muc$ (the error threshold), while it consists of random sequences (it is delocalized) for larger values. Because the error threshold $\muc \sim 1/L$ is usually inversely proportional to sequence length, the problem arises whether the maximally sustainable complexity of a self-replicator suffices to perform the complex task of self-replication~\citep{EigenNaturwissenschaften:78a}.

In a prebiotic context, it is important to emphasize that using a fitness landscape where rates depend only on the template sequence pertains to non-enzymatic rather than enzymatic replication, because in the latter case the replication rates also depend on the concentrations and the characteristics of involved enzymes. For RNA, the potential for non-enzymatic replication is questionable, given that template-directed polymerization or ligation seems limited to short molecules with rather specific sequences~\citep{KiedrowskiAC:86,AcevedoJMB:87,WuJACS:92,OrgelCRBM:04}. Moreover, experimental and simulation studies demonstrate a strongly disadvantageous tendency for elongating side-reactions at the cost of replication~\citep{FernandoJME:07}. Hence, although there have been speculations~\citep{PaceOLEB:85}, it remains unclear how a single more complex RNA should literally copy itself~\citep{JoyceAC:07,SzostakNature:01}. 

Enzymatic replication is more plausible~\citep{OrgelNature:92}, but raises the question whether high replication efficiency (high fitness) is a property of the substrate or the enzyme. In the latter case, a superior replicase does not enjoy a selective advantage, because it replicates non-functional mutant templates just as well as itself, while it is not guaranteed in the former case that a superior template is functional at all. Likewise, mutations generate substrates that are replicated less efficiently, but they also produce less-efficient enzymes~\citep{SmithNature:79}, thus affecting the replication rates of all potential substrates. On theoretical grounds, one should expect that a superior replicator is a good enzyme and a good substrate \emph{at the same time}. As enzyme, it should therefore replicate only functional substrates, and as substrate, it should be replicated preferably by efficient enzymes. It has long been realized that in a competitive environment these propensities are crucial for the emergence, improvement and perpetuation of replicase activity, which is essentially an altruistic trait that is not by itself selected for~\citep{MichodAZ:83}. The commonly proposed solution to render enzymatic replication evolutionarily stable is to impose a form of group selection, e.g., via compartmentalization in vesicles, in order to keep similar molecules closely together~\citep{SzathmaryJTB:87,AlvesPRE:01,FontanariJTB:06}, even though this requires a simultaneous and coordinated emergence of replicators and protocells~\citep{SzostakNature:01}. However, known ribozymes act with moderately or even strongly substrate-specific efficiency~\citep{JoyceAC:07}, such that \emph{specific recognition} may also have a significant influence. Since unspecific reactions require sophisticated substrate-binding properties that could well have been a rather late invention in prebiotic evolution~\citep{JohnstonScience:01}, it seems natural to assume that replication efficiency should depend strongly on the interaction between enzyme and substrate. 

In this paper, we analyze a general model of enzymatic replication accounting both for varying degrees of specificity and the characteristically broad quasi\-species distributions in order to address the consequences for the error threshold. Similar to models for the evolution of regulatory DNA motifs~\citep{GerlandJME:02}, we assume that specificity depends on the quality of binding to some recognition or tag sites~\citep{WeinerPNAS:87}. Idealizing this condition, we use replication rates that depend on the Hamming distance between these sequence regions of enzyme and substrate via a decreasing (self-specific) or increasing (cross-specific) function. After formulating the model and discussing our analytical approach in a methods section, we show for these two scenarios results from stochastic simulations and numerical solutions of deterministic rate equations.  Using our analytical toolbox, we discuss the resulting localization conditions, error thresholds and the phase diagram. In our conclusions, we use experimental values for polymerization rates to obtain simple estimates for the maximum number of nucleotides that can be used for recognition.

\section{\label{sec:model}Model}

In the framework of quasispecies theory, each molecule is characterized by its sequence $S_i = (\sigma_1^{(i)}\ldots\sigma_L^{(i)})$ of $L$ binary nucleotides $\sigma_\ell^{(i)}\in \{0,1\}$. In an infinitely large population, its concentration $X_i$ evolves according to the deterministic rate equations~\citep{EigenACP:89}
\begin{equation}\label{eq:rate-equations-general}
\dot X_i = \sum_j M_{ij} R_j X_j - X_i \sum_j R_j X_j.
\end{equation}
Here, $M_{ij}=\mu^{d_{ij}}(1-\mu)^{L-d_{ij}}$ is the mutation probability between sequences $S_i$ and $S_j$ with Hamming distance $d_{ij} = \sum_\ell |\sigma_\ell^{(i)}-\sigma_\ell^{(j)}|$, where $\mu$ is the error probability per single nucleotide (usually called ``mutation rate''), and the replication rate $R_i$ of sequence $S_i$ is given by:
\begin{equation}\label{eq:replication-rate-ansatz}
R_i = A_i + \sum_j B_{ij} X_j.
\end{equation}
Whereas the non-enzymatic rate $A_i$ depends only on the genotype $S_i$, the second term implies frequency-dependent selection and makes our model intrinsically nonlinear. It encodes the catalytic interactions of two molecules: $B_{ij}$ measures how well $S_j$ catalyzes the replication of $S_i$. The second term in Eq.~\eqref{eq:rate-equations-general} ensures the normalization $\sum_j X_j=1$, and a degradation term $-D_i X_i$ therefore drops out of Eq.~\eqref{eq:rate-equations-general} since we assume that the decay rate $D_i\equiv D$ is sequence-independent for simplicity. Note that actual RNA sequences replicate via a complementary intermediate, while our idealized model assumes direct replication in a single step. It has been shown that these two approaches are essentially equivalent for the symmetric situations considered here~\citep{StadlerMB:91}.

To capture the {pertinent features} of a situation where replicase enzymes prefer to replicate themselves instead of their competitors, we assume that the quality of specific recognition influences catalytic rates more strongly than the actual genotypes of enzyme and substrate. Hence, we effectively only model the recognition regions of ribozymes, which are often clearly separated from the catalytic domains~\citep{LilleyCOSB:05}. {Of course, the proper function of the latter region is indispensable, but in our idealized model we neglect the influence of mutations: as we have shown previously in a simple model, their effect on the error threshold is largely independent from the more interesting consequences of mutations in the recognition region~\citep{ObermayerEPL:09}. The sequence length $L$ is thus restricted} to the number of nucleotides that take part in recognition. {Probably} mediated via specific base-pairing interactions~\citep{DoudnaNature:02}, the quality of recognition can be taken as function of the number of mismatches between the binding sites of enzyme and substrate, and we let the catalytic matrix $B_{ij}$ therefore depend via a \emph{specificity function} $f(d)$ only on the Hamming distance $d_{ij}$ between enzyme and substrate. Further, because rate enhancements through ribozyme catalysis can be substantial~\citep{DoudnaNature:02}, such that non-enzymatic replication rates are comparably small (if nonzero at all), we neglect their genotype dependence altogether and choose a flat fitness landscape for $A_i$:
\begin{equation}\label{eq:model-rates}
A_i\equiv \alpha, \qquad B_{ij} = \beta f(d_{ij}).
\end{equation}
Because we are interested in the stationary state, the parameter $\alpha$ (if nonzero) merely sets the time scale while $\beta$ measures the selection strength. 

We will first analyze a scenario for self-specific replication, where replication rates increase with \emph{similarity} of enzyme and substrate. {To have the degree of specificity explicitly tunable via a parameter $p$, we use the specificity function
\begin{equation}\label{eq:self-specificity}
f_\text{s}(d) = (1-d/L)^p.
\end{equation}
As a contrasting example, we will then analyze a similarly defined specificity function where replication rates increase with \emph{complementarity} between enzyme and substrate:
\begin{equation}\label{eq:cross-specificity}
f_\text{c}(d) = (d/L)^p.
\end{equation}
The numerical and analytical approach presented in the next section can readily be applied to other functional forms of the specificity function.}

\section{Methods}

{For reasonably large sequence length, the full $2^L$-dimensional system Eq.~\eqref{eq:rate-equations-general} can only be analyzed using stochastic simulations in a finite population of $N$ sequences. Here, we employ the straightforward stochastic simulation algorithm used by \cite{WilkePR:01}. At time $t$ each sequence $S_k$, present in $N_k$ copies, has a probability $p_{0,k} = N_k/\sum_i N_i(1+R_i)$ to be copied without mutations into the population at time $t+1$, and a probability $p_{\text{mut},jk} = M_{jk} R_k N_k/\sum_i N_i (1+R_i)$ to be selected and mutated into sequence $S_j$. Following initialization, our observables of interest  are measured by averaging over time after reaching a stationary state.}

{In order to derive analytical results, we exploit that the stationary states of Eq.~\eqref{eq:rate-equations-general} are localized about a particular ``master'' sequence $S_*$, which is necessary to preserve its information content. In contrast, delocalization indicates that such a sequence $S_*$ cannot be maintained due to replication errors. Since the possibility of non-trivial dynamics such as periodic orbits cannot be excluded for general replicator-mutator equations like Eq.~\eqref{eq:rate-equations-general}~\citep{StadlerBMB:95}, there might also be other reasons for the absence of localization. However, we did not find any signs of complex dynamical behavior in our simulations,} and one can easily convince oneself that the delocalized state, where all sequences have the same concentration and therefore identical replication rates, can lead to localization: since replication rates are essentially proportional to concentration, stochastic concentration fluctuations imply higher rates. {Unless hindered by excessive mutations, this} can induce a transition to a state localized about some randomly chosen master sequence, which then also has the highest replication rate (see \citet{ObermayerEPL:09} for a visualization). Such ``fixation'' events are very similar to the phenomenon of consensus formation, e.g., in language dynamics~\citep{BlytheJSM:09}. Note that our idealized replication rates do not predetermine any specific master sequence for localization. This symmetry would be broken in a full model where replication rates depend on the full genotypes of enzyme and substrate (and not just the Hamming distance between their recognition regions).

Given localization about $S_*$, we can significantly reduce the dimensionality of Eq.~\eqref{eq:rate-equations-general} by lumping all sequences $S_i$ with a Hamming distance $k$ to $S_*$ together into ``error class'' $k$. Without loss of generality, we assume that $S_*=(00\ldots0)$. This well-known procedure~\citep{SchusterBMB:88,WoodcockJTB:96} permits to formulate reduced rate equations formally equivalent to Eq.~\eqref{eq:rate-equations-general} in terms of new variables $x_k$ denoting the concentration of error class $k$ in the population:
\begin{equation}\label{eq:rate-equations-reduced}
\dot x_k = \sum_{ji} m_{kj}[a_j\delta_{ji} + b_{ji} x_i]x_j - x_k\sum_{ji} [a_j \delta_{ji} + b_{ji}x_i]x_j.
\end{equation}
{Here and in the following, we use lowercase letters for all variables in the reduced system. For our model Eq.~\eqref{eq:model-rates}, the non-enzymatic rates are given by} $a_i\equiv \alpha$. The accordingly reduced mutation matrix and the catalytic matrix depend only on the Hamming distance between pairs of sequences in different error classes (measured with respect to the master sequence), which allows us to {combinatorially} assess all possibilities for their relative distance. The total probability of distributing $0\leq k-j+2\ell\leq L$ mutations to move a sequence from error class $j$ into error class $k$ {has been derived previously as~\citep{WoodcockJTB:96}}
\begin{equation}\label{eq:mutation-matrix-reduced}
m_{kj} = \sum_\ell \binom{L-j}{k-j+\ell}\binom{j}{\ell} (1-\mu)^{L-(k-j+2\ell)}\mu^{k-j+2\ell}.
\end{equation}
The replication rate Eq.~\eqref{eq:replication-rate-ansatz} also depends on the frequency of each sequence in each error class. With the homogeneity assumption that all $\binom{L}{j}$ sequences in class $j$ are equally populated, this complication can be resolved, and the reduced matrix $b_{ij}$ reads analogously
\begin{equation}\label{eq:specificity-matrix-reduced}
b_{ij} = \beta \sum_n \binom{L-i}{j-i+n}\binom{i}{n} \frac{f(j-i+2n)}{\binom{L}{j}}.
\end{equation}
Numerical solutions to the $(L+1)$-dimensional rate equations given in Eq.~\eqref{eq:rate-equations-reduced} can now easily be found by means of standard algorithms.

{A more detailed understanding of these solutions can be obtained from the population distribution's normalized first moment $a=\ave{k}/L=\ave{\sigma}$,} which as the mean Hamming distance $\ave{k}=\sum_k k x_k$ to the master $S_*=(00\ldots0)$ characterizes the width of the distribution and measures the mean value $\ave{\sigma}$ of each sequence's binary nucleotides. Writing down an equation for the first moment of Eq.~\eqref{eq:rate-equations-reduced} requires a hierarchy of expressions for higher moments, which can be truncated by means of a moment closure technique. For the self-specific case, we assume that the stationary Hamming distance distribution is approximately binomial, $x_k \approx \binom{L}{k} a^k (1-a)^{L-k}$, because this reproduces the expected distribution in the limits $a\to 0$ (complete localization about one sequence) and $a\to 1/2$ (the delocalized state, where the binary nucleotides are random numbers). Moreover, it solves the rate equations Eq.~\eqref{eq:rate-equations-general} exactly for linear fitness landscapes without epistasis~\citep{WoodcockJTB:96} and for an extension of the quasispecies model to a game theory setting~\citep{LassigEPL:03}. With this binomial ansatz, $a$ is the population distribution's only parameter, and for our model of the replication rates Eq.~\eqref{eq:model-rates}, it obeys the equation
\begin{multline}\label{eq:binomial-approximation-self}
(1-2a)\Big\lbrace\mu L\left[\alpha+\beta S(1-2a(1-a))\right] \\
-a(1-a) (1-2\mu)\beta S'(1-2a(1-a))\Big\rbrace = 0.
\end{multline}
This equation for $a$, which is one of our main analytical results (see \ref{app:self-derivation} for a derivation), holds for \emph{any} specificity function $f(d)$, which enters via the function 
\begin{equation}\label{eq:auxiliary-self}
S(x) = \sum_k \binom{L}{k} f(k)\, x^{L-k}(1-x)^k.
\end{equation}
{An intuitive interpretation of this auxiliary function derives from recognizing that the quantity $\alpha+\beta S(1-2a(1-a))$ measures the \emph{mean replication rate} (or mean fitness) $\sum_j R_j X_j$ of the population if $a$ solves Eq.~\eqref{eq:binomial-approximation-self}. Such solutions $a(\mu)$} can be obtained explicitly only in special cases where $S(x)$ attains a simple form (see \ref{app:self-solutions}), but we can easily solve for $\mu(a)$ and {invert graphically to obtain a \emph{bifurcation diagram}:
\begin{equation}\label{eq:bif-diag-self}
\mu(a) = \left[2+\frac{\alpha+\beta S(1-2a(1-a))}{\beta a(1-a)S'(1-2a(1-a))}\right]^{-1}.
\end{equation}

For cross-specific replication with the specificity function Eq.~\eqref{eq:cross-specificity}, we expect two equivalent subpopulations localized about complementary sequences, which corresponds to a superposition of binomial distributions: $x_k\approx\frac{1}{2}\binom{L}{k}\big[a^k(1-a)^{L-k}+(1-a)^k a^{L-k}\big]$. To obtain an equation similar to Eq.~\eqref{eq:binomial-approximation-self} for their mean widths $a$, we cannot use the first moment of the reduced rate equations (it vanishes by construction, because the distribution is symmetric about $a=1/2)$, but use the second moment $\ave{\Delta k^2}=\sum_k (k-\ave{k})^2 x_k$, leading to a lengthy result explicitly given in \ref{app:cross-derivation} and a corresponding expression for the bifurcation diagram (see Eqs.~\eqref{eq:binomial-approximation-cross} and \eqref{eq:bif-diag-cross}).
}
\section{\label{sec:results}Results and Discussion}

\subsection{Self-specific replication}
\begin{figure*}[p]
\centerline{\includegraphics{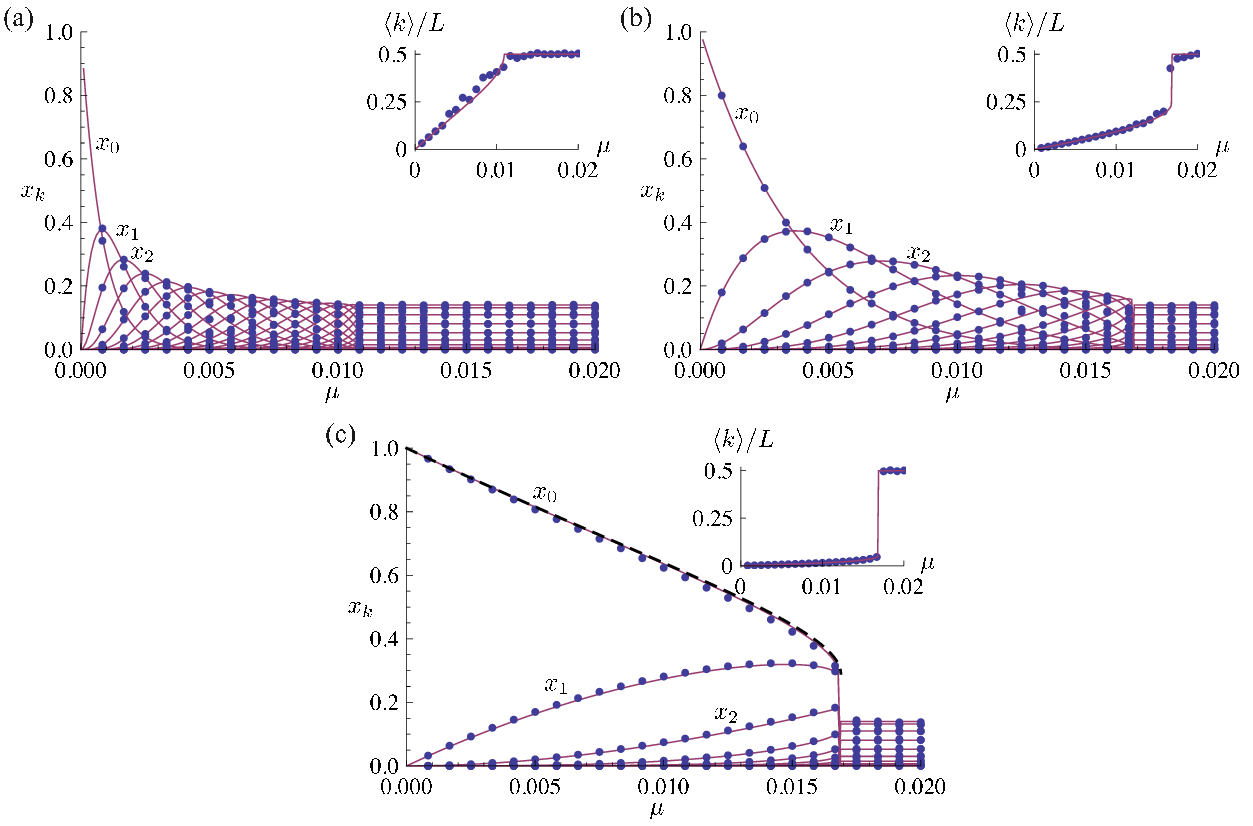}}
\caption{\label{fig:plot-egf}Solutions for self-specific replication with $f_\text{s}(d)=(1-d/L)^p$. Stationary Hamming distance distribution $x_k$ as function of mutation rate $\mu$ from numerical solutions to the reduced rate equations Eq.~\eqref{eq:rate-equations-reduced} (straight lines) and a stochastic simulation of the full system Eq.~\eqref{eq:rate-equations-general} in a population of $N=10^4$ sequences of length $L=32$ (dots) for $\alpha=1$, $\beta=5$ and (a) $p=1$, (b) $p=5$, and (c) $p=\infty$ (here, the dashed line shows the error-tail approximation for $x_0$). The insets depict the average Hamming distance $a=\ave{k}/L$ to the master sequence.}
\end{figure*}

Our first scenario is concerned with self-specific replication, where replication rates increase with similarity of enzyme and substrate through the specificity function Eq.~\eqref{eq:self-specificity}. {As argued in the preceding section, stochastic fluctuations increasing the frequency of one particular sequence also increase its replication rate, such that the population can localize about this master sequence $S_*$. Figure~\ref{fig:plot-egf} shows the stationary Hamming distance distribution $x_k$, which measures the concentration of sequences with $k$ mutations relative to $S_*$, for different degrees of specificity from the linear case $p=1$ to complete self-specificity $p\to\infty$, where enzyme and substrate have to be identical. We compare results from stochastic simulations to numerical solutions of the reduced rate equations, Eq.~\eqref{eq:rate-equations-reduced}. The excellent agreement between simulation and deterministic theory justifies the homogeneity assumption made in symmetrizing the specificity matrix (see Eq.~\eqref{eq:specificity-matrix-reduced}).  In the simulations shown in Fig.~\ref{fig:plot-egf}, we initialized all sequences at a predetermined master sequence in order to avoid noise from the intrinsically stochastic ``fixation'' events, but we also tested other initial conditions with modest inhomogeneities, which were quickly ``washed out'' and did not give rise to measurable differences in the stationary state.}

The limit $p\to\infty$ with $f_\text{s}(d)\to\delta_{d,0}$, depicted in Fig.~\ref{fig:plot-egf}(c), leads to a generalized Schl\"ogl model of auto-catalytic replication, which has been partly analyzed by \cite{StadlerBMB:95}. In this limit we can employ the well-known ``error-tail'' approximation~\citep{SchusterBMB:88}: we define $x_0$ as the concentration of the master sequence, $\alpha+\beta x_0$ its replication rate and $(1-\mu)^L$ the probability not to have a mutation. All other sequences are lumped together in the error tail with concentration $1-x_0$ and replication rate $\alpha$ (the concentration of suitable {replicase} enzymes is so small that the frequency-dependent term in the replication rate does not contribute). Neglecting back mutations from the error tail into $x_0$ (corresponding to the large-genome limit), we obtain the simple equation
\begin{equation}\label{eq:rate-equation-error-tail}
\dot x_0 = (\alpha+\beta x_0)(1-\mu)^Lx_0 - x_0 \bar r ,
\end{equation}
with $\bar r = (\alpha + \beta x_0)x_0+\alpha(1-x_0)$ the mean replication rate. In the stationary state, we easily find that the delocalized state $x_0=0$ is stable for all $\mu$, while a branch of solutions with nonzero $x_0$ emerges for $(1-\mu)^L > 2(\sqrt{\alpha(\alpha+\beta)}-\alpha)/\beta$ through a \emph{discontinuous} transition (see also \citet{CamposPRE:00,ObermayerEPL:09,WagnerPRL:10} for similar results in related models). Whereas for the somewhat related sharply-peaked fitness landscape, where only the master sequence has a higher replication rate, the error threshold arises through a \emph{continuous} bifurcation~\citep{BaakeJMB:97}, the discontinuity observed here expresses the qualitatively different behavior we previously termed ``escalation of error catastrophe''~\citep{ObermayerEPL:09}: as the mutation rate grows, the proportion of fittest sequences, i.e., of the necessary {replicase} enzymes, is diminished and therefore their replication rate. This in turn reduces their concentration, until at the error threshold the concentration of enzymes $x_0$ is not large enough to have them replicate with an efficiency sufficient for localization. 

\begin{figure}[p]
\centerline{\includegraphics{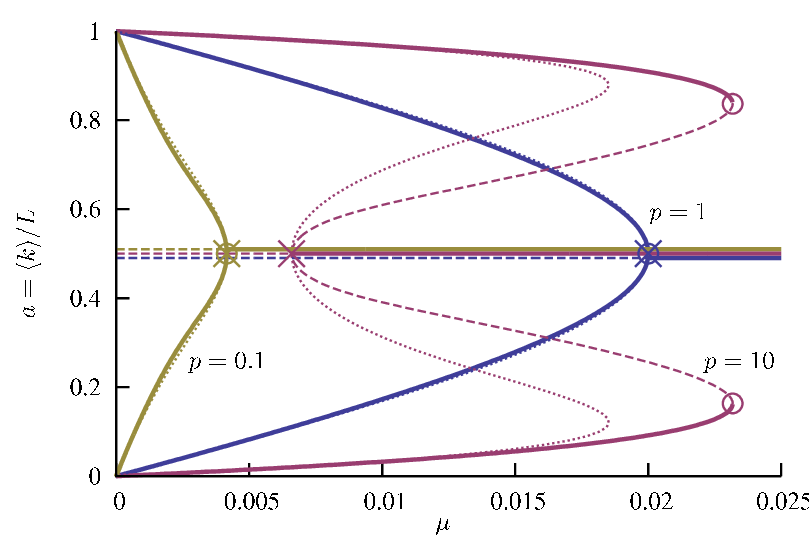}}
\caption{\label{fig:bif-diag-self}Bifurcation diagram of Eq.~\eqref{eq:rate-equations-reduced} for the normalized mean Hamming distance $a=\ave{k}/L$ with $L=8$, $\alpha=1$, $\beta=1$ and different values of the specificity degree $p$. Thick lines indicate stable branches, dashed lines unstable branches and thin dotted lines the results of the binomial closure approximation Eq.~\eqref{eq:binomial-approximation-self}, which is barely visible for $p=0.1$ and $p=1$. Circles indicate critical mutation rates $\muc$ where the localized regime vanishes, crosses show values $\muct$ where the delocalized state changes stability (horizontal branches corresponding to delocalized states are drawn slightly shifted for visualization). }
\end{figure}

Although Eq.~\eqref{eq:rate-equation-error-tail} approximates the exact result for $x_0$ very well (see the dashed line in Fig.~\ref{fig:plot-egf}(c)), it is valid only for $L\to\infty$ (because back mutations are neglected) and $p\to\infty$ (because the replication rate of the error-tail is taken as concentration-independent). {In order to gain a more general perspective, we use the analytical solutions of Eq.~\eqref{eq:binomial-approximation-self} for the mean Hamming distance to the master (the population distribution's first moment $a=\ave{k}/L$) obtained from the bifurcation diagram Eq.~\eqref{eq:bif-diag-self}.} A comparison between the exact solution $a(\mu)$ resulting from the reduced rate equations Eq.~\eqref{eq:rate-equations-reduced} via numerical continuation\footnote{AUTO software package available via \tt{http://indy.cs.concordia.ca/auto/.}} and Eq.~\eqref{eq:bif-diag-self} is shown in Fig.~\ref{fig:bif-diag-self} for $L=8$, $\alpha=\beta=1$ and different values of $p$ in the specificity function Eq.~\eqref{eq:self-specificity}. Recalling the symmetry of the original model, Eq.~\eqref{eq:model-rates}, namely that the population can localize about any sequence, the remaining reflection symmetry $a\to 1-a$ about the delocalized solution $a=1/2$ indicates that after symmetrization localization is only possible about the master sequence or its complement. The stable branches associated with localized solutions start for $\mu=0$ at $a=0$ (or $a=1$), i.e., full localization about one sequence (or its complement), and higher mutation rates give rise to broader distributions with larger mean, until these localized regimes disappear at critical mutation rates $\muc$ denoted by circles. The delocalized solutions, on the other hand, gain stability at finite mutation rates $\muct$ (denoted by crosses), and we find bistability for $\muct<\mu<\muc$. {A similar situation is encountered in formally related models of grammar evolution~\citep{NowakScience:01,KomarovaJTB:04}.}

Explicit expressions for the error threshold $\muc=\max \mu(a)$ {are available from Eq.~\eqref{eq:bif-diag-self} when the auxiliary function $S(x)$ (defined in Eq.~\eqref{eq:auxiliary-self})} has a simple form (see \ref{app:self-solutions}). For instance, $\muc=\beta/(4 \alpha L +2\beta (L+1))$ for $p=1$, where specificity decreases linearly with the distance between enzyme and substrate. Although an error threshold is absent for linear fitness landscapes without epistasis~\citep{WoodcockJTB:96}, here the intrinsically nonlinear model gives a {sharp} transition even in this apparently similar case. For very strong specificity $p\to\infty$ we find $\muc=\mathcal{W}[\beta/(\e\alpha)]/(2L)$ using Lambert's $\mathcal{W}$-function, and we recover the result $\muc=\ln(\beta/\alpha)/(2L)$ that can also be obtained from the error-tail approximation in the limit $\beta\gg \alpha$.  Further, the delocalized state $a=1/2$ is the only solution of Eq.~\eqref{eq:binomial-approximation-self} in the complete absence of specificity ($f_\text{s}(d)\equiv 1$), supporting the intuition that unspecific replication does not suffice to preferentially maintain the information of one particular sequence. Interestingly, taking $p\to 0$ in Eq.~\eqref{eq:self-specificity} gives the finite even though exponentially small value $\muc=\beta/(2^{L+1}(\alpha+\beta))$.
This result implies that limited localization is possible even for very weak specificity (if $p=0$ in the specificity function Eq.~\eqref{eq:self-specificity}, enzymes replicate everything except their exact complement, because always $f_\text{s}(L)=0$).

From the bifurcation diagram obtained via Eq.~\eqref{eq:bif-diag-self}, we easily read off exact results for the value $\muct=\mu(1/2)$ where the delocalized state gains stability. E.g., for the generalized Schl\"ogl model $p\to\infty$, we get $\muct = \beta/(\alpha 2^{L+1}+4\beta)$~\citep{StadlerBMB:95}. This exponentially small yet finite value is consistent with our previous conclusion that the delocalized regime is stable for all values of $\mu$ within the error-tail approximation, Eq.~\eqref{eq:rate-equation-error-tail}, which holds for $L\to\infty$.  Further, we find that the two critical values $\muct$ and $\muc$ are identical for $p=0,1,2$. Recognizing that Fig.~\ref{fig:bif-diag-self} describes a pitchfork bifurcation at $a=1/2$ and $\mu=\muct$, we infer that the two critical mutation rates are equal ($\muc=\muct$) whenever the pitchfork is supercritical, whereas bistability between localized and delocalized states for intermediate mutation rates $\muct < \mu < \muc$ is possible in the subcritical case, leading to the discontinuous transition observed in Fig.~\ref{fig:plot-egf}. The bistability regime vanishes as the curvature $\mu''(1/2)$ in the bifurcation diagram changes sign, which gives from  Eq.~\eqref{eq:bif-diag-self} an approximate expression for the corresponding critical value of $\beta$:
\begin{equation}\label{eq:beta-star-self}
\beta^* = \alpha \left[\frac{S'^2(1/2)}{S''(1/2)-2 S'(1/2)} - S(1/2)\right]^{-1}.
\end{equation}
{This general result can readily be evaluated for any specificity function entering the auxiliary function $S(x)$ defined in Eq.~\eqref{eq:auxiliary-self}. It predicts bistability for all values $\beta < \beta^*= \alpha\,2^{L-1} (L-2)$ if $p\to\infty$ in our choice Eq.~\eqref{eq:self-specificity} of the specificity function, and no bistability for weak specificity because the critical value of the coupling constant is negative ($\beta^* \leq 0$) for small $p < p_\text{min} = 2+\mathcal{O}(L^{-1})$. Since the mean Hamming distance $\ave{k}=a L$ is discontinuous at the error threshold only in the bistability regime $\beta < \beta^*(p)$, we argue that the observation of a discontinuous mean replication rate (mean fitness) $\alpha+\beta S(1-2a(1-a))$ depends on the specificity of second order catalysis, which generalizes the result of~\citet{WagnerPRL:10}.}

\begin{figure}[p]
\centerline{\includegraphics{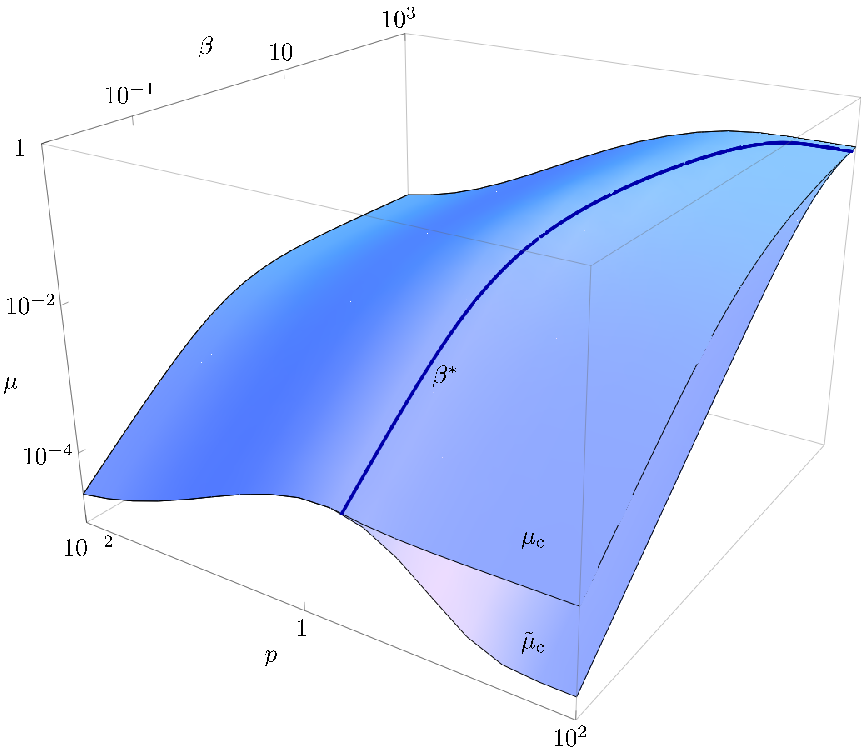}}
\caption{\label{fig:phasediagram-self} Phase diagram of localization regimes in the parameter space of mutation rate $\mu$, selection strength $\beta$ and specificity degree $p$ (log-log-log scale) obtained from Eq.~\eqref{eq:binomial-approximation-self} for $L=8$ and $\alpha=1$: below the upper plane $\muc$, a localized solution exists, while above the lower plane $\muct$ the delocalized state becomes stable. Bistability ($\muc > \muct$) is possible only for $\beta < \beta^*(p)$.}
\end{figure}

\begin{figure*}[p]
\centerline{\includegraphics{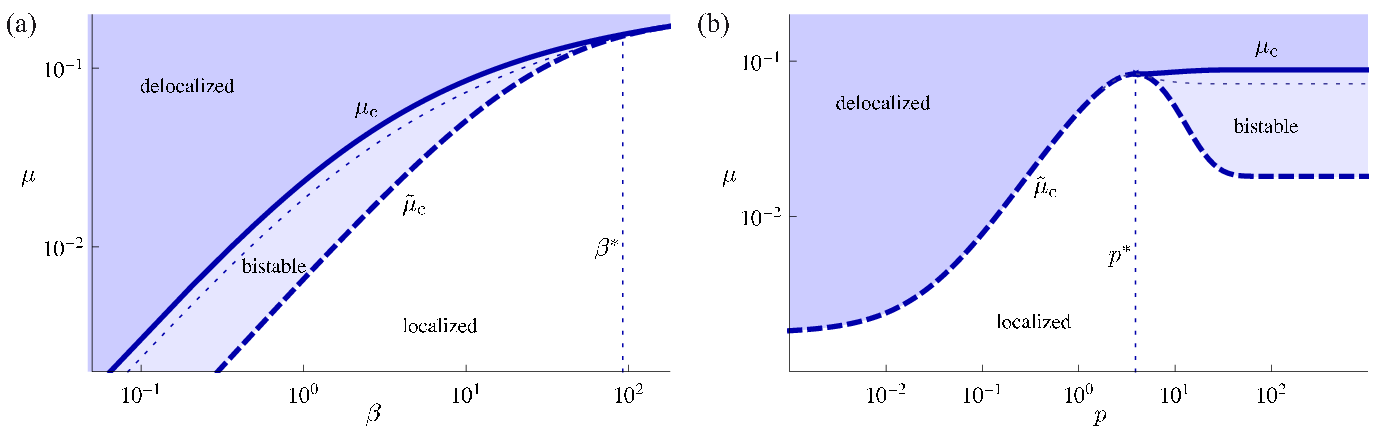}}
\caption{\label{fig:phasediagram-self-bp}Projected phase diagrams of localization regimes for self-specific replication with $L=8$ and $\alpha=1$ {obtained from Eq.~\eqref{eq:rate-equations-reduced}}: (a) as function of $\beta$ with $p=10$ fixed; (b) as function of $p$ with $\beta=10$ fixed. Below  $\muc$ (thick line), a localized solution exists, while above $\muct$ (thick dashed line) the delocalized state becomes stable. The dotted lines denote the result for $\muc$ obtained via the binomial closure approximation. Bistability ($\muc > \muct$) is possible only for $\beta < \beta^*(p)$ (or $p>p^*(\beta)$).}
\end{figure*}

The results of the binomial closure approximation Eq.~\eqref{eq:binomial-approximation-self} are summarized in the phase diagram Fig.~\ref{fig:phasediagram-self}, where the two critical mutation rates $\muc$ and $\muct$ are shown as functions of the selection strength $\beta$ and the specificity degree $p$ for $\alpha=1$. The thick line denoted $\beta^*$ {indicates} the boundary of the bistability regime $\muc > \muct$. Fig.~\ref{fig:bif-diag-self} demonstrates that the binomial approximation is quantitatively excellent in the supercritical situation $\beta > \beta^*$ {(in particular, it gives exact results for $\muc$ and $\muct$)}, and qualitatively correct otherwise, where the values for $\muc$ are somewhat underestimated: near the error threshold, the variance of the population distribution is considerably larger than that of a binomial. {Nevertheless, we emphasize that the result $\muc=\ln(\beta/\alpha)/(2 L)$ in the limiting case $p\to\infty$ agrees with the value obtained from the error-tail approximation for large $\beta$. The remarkable performance} of the binomial closure approximation can be  appreciated in more detail from the projected phase diagrams shown in Fig.~\ref{fig:phasediagram-self-bp}.

A {noticeable} feature of these phase diagrams is that the error threshold $\muc$ \emph{increases} for stronger specificity $p$ (see Fig.~\ref{fig:phasediagram-self-bp}(b)), which implies that higher mutation rates can be tolerated, i.e., that longer sequences can be maintained. This seems at first counter-intuitive, because weaker specificity constraints on the recognition sequence should allow more mutational ``freedom''. However, as shown in Fig.~\ref{fig:bif-diag-self} and Fig.~\ref{fig:plot-egf}, smaller values for $p$ lead to much broader distributions: mutants are still reasonably well replicated by master enzymes, but the master is only moderately well (but not quite as efficiently) replicated by the mutants. The resulting broadening of the distribution effectively reduces the replication rate of the master and escalates the error catastrophe~\citep{ObermayerEPL:09}. Thus, the necessity for an enzymatic replicator to discriminate not only between functional and non-functional \emph{substrates}, but also between efficient and unproductive \emph{enzymes}, is again emphasized.

We finally want to remark on the case $\alpha=0$, i.e., no background level for the non-enzymatic replication rate.  Most of the above results obtained from the binomial closure approximation can be simply evaluated for $\alpha=0$ (note that then $\beta$ sets the timescale and drops out), but the strong specificity limit $p\to\infty$ deserves extra attention. The error-tail approximation indicates that the error threshold vanishes ($\muc\to 1$), but from Eq.~\eqref{eq:binomial-approximation-self} we find the {\emph{exact result}} $\muc=\muct=1/6$ for  $L\gg 1$, i.e., a macroscopic yet finite value (see \ref{app:self-solutions}). This remarkable result can be explained by recalling that the traditional result $\muc\approx \ln r/L$~\citep{EigenACP:89,WieheGR:97} for the error threshold depends on the replication advantage $r$ of the master relative to a possibly small but finite value for the mutants. In our case, rates are directly proportional to concentration, and because the master sequence has a concentration of order 1, while in an infinitely large population distant mutants have concentrations of order $2^{-L}$, this relative advantage itself is of order $2^L$, and cancels the length dependence of the error threshold. In the corresponding non-enzymatic case, results for so-called ``truncation'' fitness landscapes have lead to some debate about the applicability of the error threshold concept in the presence of lethal mutations~\citep{WilkeBEB:05,SummersJV:06,TakeuchiBEB:07,SaakianPRE:09}. Accordingly, we should cautiously note that our results for $\alpha=0$ will probably be affected when accounting for the effects of finite populations and the full dependence of the replication rates on the genotypes of enzyme and substrate.

\subsection{Cross-specific replication}
To increase the information content of replicating systems beyond the limited complexity of a single replicator, auto-catalytic reaction networks such as hypercycles~\citep{EigenNaturwissenschaften:78a,StadlerBMB:95} have been proposed, where different molecular species catalyze each other's replication in a possibly complex interaction graph. Only very little is known for these systems regarding the issue of reaction specificity and the cross-interactions of each species' mutant clouds. The simplest conceivable networks are 2-member cross-catalytic hypercycles. {While the two members (which actually replicate via complementary intermediates) need not be strictly complementary, we assume that they are sufficiently distinct that an idealized model requiring complementary recognition regions captures the essentials.} This suggests to analyze the specificity function Eq.~\eqref{eq:cross-specificity} where replication rates increase with Hamming distance between enzyme and substrate. {In this case}, we expect the formation of two sub-populations localized about complementary sequences, each catalyzing the replication of the other. The main question to be answered is how specificity affects coexistence. In the standard Eigen model, quasispecies coexistence is prevented by competitive exclusion except in degenerate cases, because the ``fittest'' individuals take over the population~\citep{SwetinaBC:82}. {Interactions between sub-populations, e.g., based on complementarity of binary traits~\citep{OliveiraPRL:02}, are known to enable coexistence. In our case, we expect novel conditions for coexistence: even though each subpopulation depends on the presence of the other for efficient replication, it is unclear how the possibly broad mutant distributions influence each other.}

\begin{figure}[p]
\centerline{\includegraphics{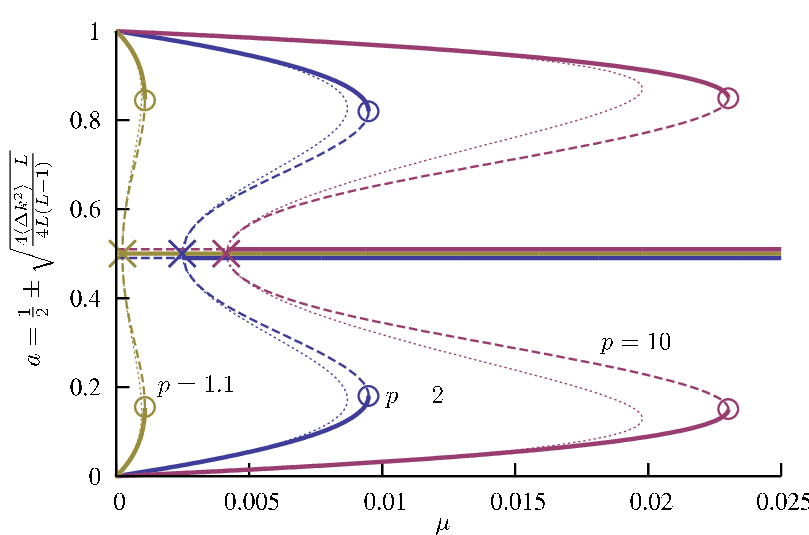}}
\caption{\label{fig:bif-diag-cross}Bifurcation diagram of Eq.~\eqref{eq:rate-equations-reduced} as in Fig.~\ref{fig:bif-diag-self}, but for cross-specific replication in terms of the population parameter $a$, which denotes the width of the two subpopulations and is calculated from the variance $\ave{\Delta k^2}$ in the Hamming distance distribution. Parameters are $L=8$, $\alpha=1$, and $\beta=1$.}
\end{figure}

\begin{figure*}[p]
\centerline{\includegraphics{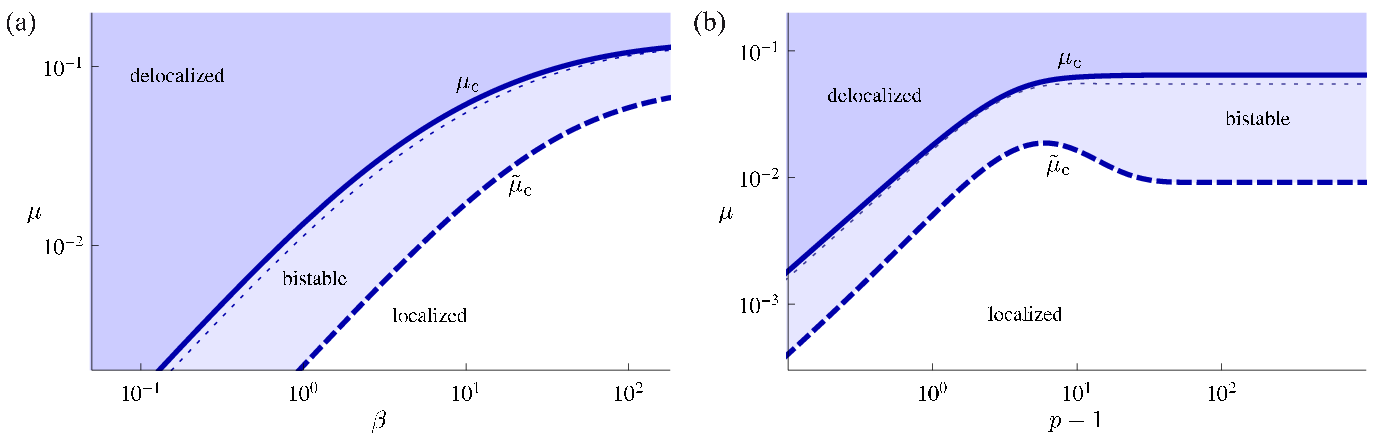}}
\caption{\label{fig:phasediagram-cross-bp}Projected phase diagrams of localization regimes as in Fig.~\ref{fig:phasediagram-self-bp}, but for cross-specific replication with $L=8$ and $\alpha=1$: (a) as function of $\beta$ with $p=10$ fixed; (b) as function of $p$ with $\beta=10$ fixed.}
\end{figure*}

From numerical solutions to the reduced rate equations and simulation results, where we initialized the population split between the master sequence and its complement (see Fig.~\ref{fig:plot-ecf}), we find that localization is only possible for $p > 1$. {This important result is confirmed through the binomial moment closure approximation  (see \ref{app:cross-bifurcation}), which has for $p=1$ only the solution $a=1/2$, corresponding to the delocalized state. Because the complementary distributions overlap too much if catalytic rates increase only linearly with Hamming distance, the populations are not localized strongly enough to ensure coexistence.

\begin{figure*}[p]
\centerline{\includegraphics{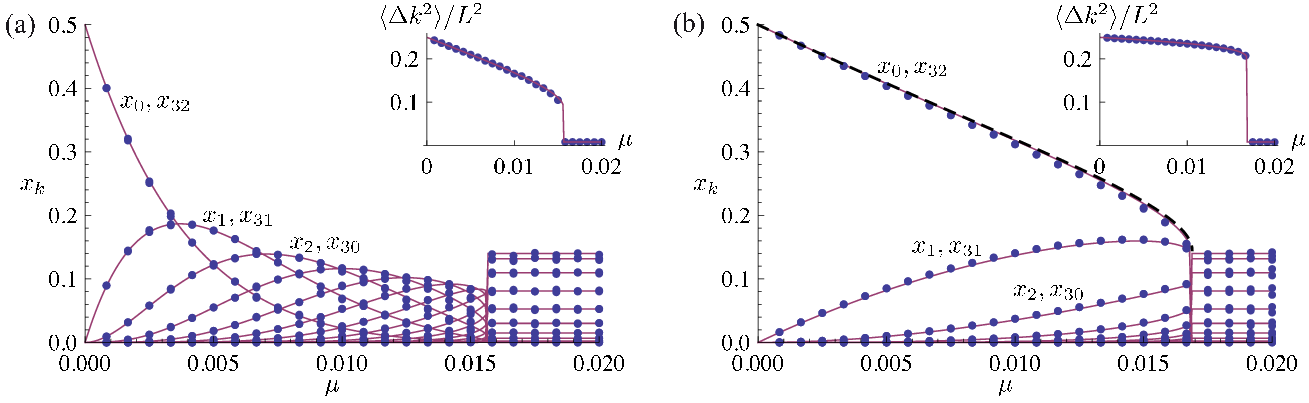}}
\caption{\label{fig:plot-ecf}Hamming distance distribution $x_k$ as function of mutation rate $\mu$ as in Fig.~\ref{fig:plot-egf}, but for cross-specific replication with $\alpha=1$, $\beta=10$, $L=32$, and (a) $p=5$ or (b) $p=\infty$. The insets show the variance $\ave{\Delta k^2}=\sum_k (k-\ave{k})^2 x_k$ (in a symmetric population, {$\ave{k}=L/2$}).}
\end{figure*}

The bifurcation diagram $\mu(a)$ obtained from our approximation scheme (see Eq.~\eqref{eq:bif-diag-cross}) is shown in comparison with the exact result from the reduced rate equations} in Fig.~\ref{fig:bif-diag-cross} for different values of $p$. We obtain information about {the parameter $a$} from the variance  $\ave{\Delta k^2}=\sum_k (k-\ave{k})^2 x_k$, in the case of two complementary binomials given by $\ave{\Delta k^2} = L^2/4-L(L-1)a(1-a)$. Using this expression {to infer $a$ from the measured variance} gives very good agreement between binomial closure approximation and exact numerical results especially for small $p$, because now the first \emph{two} moments are correct. The critical mutation rates $\muc$ and $\muct$ can be found from the bifurcation diagram, e.g., $\muc=\mathcal{W}[\beta/(2\alpha\e)]/(2 L)$ for $p\to\infty$, which is identical to the corresponding result for self-specific replication if we replace $\beta\to\beta/2$. Hence, in this limit we obtain two clearly separated binomial distributions representing two equivalent and catalytically coupled populations: {indeed, in the localized state the sum $x_k+x_{L-k}$ from Fig.~\ref{fig:plot-ecf}(b) is equal to $x_k$ in the self-specific situation shown in Fig.~\ref{fig:plot-egf}(c) once we replace $\beta\to\beta/2$. The} ``coupling constant'' is only half as large because only one half of the population is available as enzymes for the other. In particular, the negligible interaction between the respective mutant clouds allows one to employ the error-tail approximation assuming independent species and error tails as in \citep{CamposPRE:00,ObermayerEPL:09}. Further, we find that the pitchfork bifurcation is always subcritical, i.e., that $\muc > \muct$.  We summarize our main results by plotting projected phase diagrams of $\muc$ and $\muct$ as functions of $p-1$ and $\beta$ in Fig.~\ref{fig:phasediagram-cross-bp}. This confirms that both critical mutation rates $\muc$ and $\muct$ vanish linearly with $p-1$, because $\mu(a)\propto p-1$ as $p\to 1$, which gives the sharp bound $p > 1$ for coexistence of two populations. Finally, the case $\alpha=0$ of zero non-enzymatic replication rate is similar {to} the self-specific case: the slight chance of distant mutants to find an appropriate enzyme gives an enormous replication advantage to the mainly populated master sequences and therefore macroscopic values for the error threshold (see \ref{app:cross-solutions} for details).

\section{Conclusion} 

Because enzymatic replication provokes the necessity for enzymes to favor functional substrates and for substrates to prefer efficient enzymes, we analyzed {the effects of specific replication on the error threshold. In our model, which accounts for the full quasispecies mutant distributions, replication rates depend on the Hamming distance between enzyme and substrate via an adjustable degree of specificity.} Combining stochastic simulations, numerical solutions of reduced rate equations and analytical solutions to a binomial closure approximation, we could analyze the entire phase diagram and assess how mutation rate $\mu$, selection strength $\beta$ and specificity degree $p$ influence the localization about a master sequence in order to preserve its information content. We found that for self-specific replication very weak specificity suffices for localization, whereas stronger specificity gives more tolerance against mutations but leads to bistability with the delocalized regime of random sequences. In particular, the binomial closure approximation permits to obtain analytical expressions for the bifurcation diagram and an upper bound $\beta^*$ for the bistability regime, which can be evaluated for any specificity function $f_\text{s}(d)$. Apart from our special choice, Eq.~\eqref{eq:self-specificity}, a mesa-shaped function would also be conceivable, in correspondence to fitness landscapes for transcription factor binding allowing for some ``fuzziness'' or neutrality in the binding sequence~\citep{GerlandJME:02}. Preliminary results indicate that in this case the binomial closure approximation gives at least qualitative agreement as well. While our approximation is not restricted to large $L$, this limit can probably be more systematically be described using the maximum principle employed previously in quasispecies theory~\citep{HermissonTPB:02,SaakianPNAS:06}. In the case of cross-specificity, we found that coexistence of subpopulations localized about complementary sequences is possible only if replication rates increase faster than linearly with Hamming distance. 

Although our model is based on idealizing assumptions, we can extrapolate from currently available experimental data to obtain rough quantitative estimates for the maximal length $L_\text{c}$ of the recognition or tag sequences that can be used by {replicase} enzymes to specifically and reliably discriminate appropriate and useless templates (and vice versa). Considering that non-enzymatic template-directed polymerization rates are on the order of several hours to days per base~\citep{AcevedoJMB:87,WuJACS:92}, while ribozyme-catalyzed polymerization gives rates in the hour range~\citep{JohnstonScience:01, ZaherRNA:07}, we can estimate the ratio $\beta/\alpha$ somewhere near $5$-$20$ if polymerization is the rate-limiting step. Assuming a self-specific enzymatic replicator with a mutation rate on the order of 3\% as in \citep{JohnstonScience:01}, we obtain a critical length $L_\text{c}=11$-$15$ for weak specificity $p=1$, and a larger value $L_\text{c}=18$-$33$ for $p\to\infty$, because stronger specificity constraints allow longer sequences due to better localization. These values are significantly smaller than the lengths of, e.g., the 154-nucleotide specificity domain of \emph{Bacillus subtilis} RNase~P~\citep{LilleyCOSB:05} or the tRNA-like structures supposed to act as ``genomic tags'' for the replication of RNA viruses~\citep{WeinerPNAS:87}. Many of the nucleotides in these instances have a structural role, which makes them effectively redundant or neutral{~\citep{KunNG:05}}, and only a minority is actually involved in recognition. Also, recent research indicates that ``stalling'' of polymerization after mismatch incorporation might significantly reduce the error threshold{~\citep{RajamaniJACS:10}}. Nevertheless, our result suggests that the error threshold puts hard constraints on the information content of enzymatic replicators as well.

\section*{Acknowledgements}
We gratefully acknowledge helpful discussions with Joachim Krug and Irene Chen and financial support by the German Excellence Initiative via the program ``Nanosystems Initiative Munich'' and by the Deutsche Forschungsgemeinschaft through SFB TR12.

\appendix

\section{\label{app:self-specific}Self-specific replication}
\subsection{\label{app:self-derivation}Derivation of Eq.~\eqref{eq:binomial-approximation-self}}

To obtain Eq.~\eqref{eq:binomial-approximation-self}, we compute the first moment $\sum_k k \dot x_k$ of the reduced rate equations Eq.~\eqref{eq:rate-equations-reduced} under the assumption that $x_k =x_k^\text{b} \equiv \binom{L}{k} a^k (1-a)^{L-k}$ is binomially distributed. This gives four terms
\begin{equation}\label{eq:T1T4}
\begin{aligned}
T_1 &= \sum_{jk} k\, m_{kj} a_j x_j^\text{b} &\qquad
T_2 &= \sum_{ijk} k\, m_{kj} b_{ji} x_i^\text{b} x_j^\text{b} \\
T_3 &= -\sum_k k\, x_k^\text{b} \sum_j a_j x^\text{b}_j &\qquad
T_4 &= - \sum_k k\, x_k^\text{b} \sum_{ij} b_{ji} x_i^\text{b} x_j^\text{b}.
\end{aligned}
\end{equation}
The reduced mutation matrix $m_{kj}$ and the catalytic matrix $b_{ij}$ are given in Eqs.~\eqref{eq:mutation-matrix-reduced} and \eqref{eq:specificity-matrix-reduced}. Note that although not immediately obvious, the catalytic matrix $b_{ij}=b_{ji}$ is symmetric, because the binomial in the denominator of Eq.~\eqref{eq:specificity-matrix-reduced} normalizes it to the single sequence level. We further keep in mind that $\binom{n}{k}=0$ if $k<0$ or $k>n$ if $n$ and $k$ are integer, so we do not need to keep track of the summation limits in the following calculations.

While it is straightforward to find $T_1=\alpha L (a+\mu(1-2a))$ and $T_3 = -\alpha a L$, we concentrate first on $T_4$, which reads after performing the summation over $k$ 
\begin{equation}
T_4 = -\beta a L\sum_{ijn}\binom{L-j}{i-j+n}\binom{L}{j}\binom{j}{n} f_\text{s}(i-j+2n) a^{i+j}(1-a)^{2L-(i+j)}.
\end{equation}
Replacing $i'=i-j+2n$ and rearranging $\binom{L-j}{i'-n}\binom{L}{j}\binom{j}{n}=\binom{L-i'}{j-n}\binom{L}{i'}\binom{i'}{n}$, we can sum over $j$ and $n$, and are left with
\begin{equation}
\begin{split}
T_4 &= -\beta a L \sum_{i'} \binom{L}{i'} f_\text{s}(i') (1-2a(1-a))^{L-i'}(2a(1-a))^{i'}\\ &= -\beta a L S(1-2a(1-a)).
\end{split}
\end{equation}
The term $T_2$ can after similar rearrangements be written as the product of two generalized Vandermonde matrices:
\begin{equation}
T_2 = \beta \sum_{i'} \binom{L}{i'} f_\text{s}(i') \sum_{jk} k \begin{bmatrix} \mu & 1-\mu \\ 1-\mu & \mu \end{bmatrix}_{L-k,j}
\begin{bmatrix} (1-a)^2 & a(1-a) \\ a^2 & a(1-a) \end{bmatrix}_{j,i},
\end{equation}
where the $L^\text{th}$ Vandermonde matrix with parameters $a$, $b$,
$c$, and $d$ is defined as
\begin{equation}\label{eq:def-vandermonde}
  \begin{bmatrix} a & b\\ c & d\end{bmatrix}_{i,j}
    \equiv\sum_\ell\binom{L-j}{i-\ell}\binom{j}{\ell}a^{L+\ell-i-j}b^{j-\ell}c^{i-\ell}d^\ell.
\end{equation}
This allows us to  use a nice multiplication identity~\citep{RawlingsMM:05} for these matrices:
\begin{equation}
  \begin{bmatrix}a & b\\ c & d\end{bmatrix}
    \begin{bmatrix} e & f \\g & h \end{bmatrix} =
    \begin{bmatrix} ae+bg & af+bh \\ ce+dg & cf+dh\end{bmatrix},
\end{equation}
which gives:
\begin{align}
T_2 =& \beta \sum_{i'} \binom{L}{i'} f_\text{s}(i') \sum_k k 
\begin{bmatrix} a^2+\mu(1-2a) & a(1-a) \\ (1-a)^2-\mu(1-2a) & a(1-a) \end{bmatrix}_{L-k,i'} \\
=& \beta\big[L(a+\mu(1-2a))S(1-2a(1-a))\\& -(1-2a)a(1-a)(1-2\mu) S'(1-2a(1-a))\big].
\end{align}
Adding up $T_1+T_2+T_3+T_4=0$ gives Eq.~\eqref{eq:binomial-approximation-self}.

\subsection{\label{app:self-solutions}Solutions of Eq.~\eqref{eq:binomial-approximation-self}}

This equation can be solved whenever $S(x)$, defined in Eq.~\eqref{eq:auxiliary-self}, assumes a simple form. For $f_\text{s}(d)=(1-d/L)^p$ as in Eq.~\eqref{eq:self-specificity}, $S(x)$ is a polynomial of order $L$, except for integer $0 < p < L$, where it is of order $p$. A few instances for $L > 2$ are given by 
\begin{equation}\label{eq:instances-S}
S(x) =\begin{cases}
1-(1-x)^L,&p=0, \\
x,&p=1, \\
\frac{1}{L} x + \frac{L-1}{L} x^2,&p=2,\\
x^L,&p=\infty.\end{cases}
\end{equation}

To find solutions $a(\mu)\neq 1/2$, we write $x=a(1-a)$ and solve Eq.~\eqref{eq:binomial-approximation-self} for $x$:
\begin{equation}
x=
\begin{cases}
\frac{1}{2}\left(2\mu\frac{\alpha+\beta}{\beta}\right)^{1/L},&p=0 \\
\frac{\alpha+\beta}{\beta}\frac{\mu L}{1+2\mu(L-1)},&p=1\\
\mu-\frac{1}{2L}\mathcal{W}\left(-\frac{2\alpha \mu L}{\beta}\e^{2\mu L}\right),&p=\infty.
\end{cases}
\end{equation}
For the last case, we approximated $(1-2x)^L\approx \e^{-2x L}$ and $\mu/(1-2\mu)\approx \mu$ for the important asymptote $L\gg 1$ with $\mu L$ fixed, and used Lambert's $\mathcal{W}$-function. A more complicated expression is obtained for $p=2$. The solution $a=\frac{1}{2}\left(1\pm\sqrt{1-4x}\right)$ is then easily computed, and the error thresholds $\muc$ follow from evaluating the condition $x = 1/4$ (for $p=0,1,2$), or from requiring the $\mathcal{W}$-function to give a real result (for $p=\infty$):
\begin{equation}
  \muc = \begin{cases}
\frac{\beta}{\alpha+\beta}2^{-(L+1)},&p=0,\\
\frac{\beta}{4\alpha L+2\beta(L+1)},&p=1,\\
\frac{\beta}{4\alpha L+\beta(L+3)},&p=2,\\
\frac{1}{2L}\mathcal{W}\left(\frac{\beta}{\e\alpha}\right), &p=\infty.
\end{cases}
\end{equation}

Finally, it is easy to evaluate the bifurcation diagram {Eq.~\eqref{eq:bif-diag-self}} at $a=1/2$ to get the critical mutation rate $\muct=\mu(1/2)$:
\begin{equation}
\muct = \frac{\beta S'(1/2)}{2\beta S'(1/2) + 4 L (\alpha+\beta S(1/2))}.
\end{equation}
For our specificity function $f_\text{s}(d)=(1-d/L)^p$, we obtain explicitly
\begin{equation}
\muct = \begin{cases}
2^{-(L+1)}\frac{\beta}{\alpha+\beta},&\quad p=0\\
\frac{\beta}{4 \alpha L +2\beta(L+1)},&\quad p=1\\
\frac{\beta}{4 \alpha L+ \beta(L+3)},&\quad p=2\\
\frac{\beta}{\alpha 2^{L+1}+4 \beta},&\quad p=\infty.
\end{cases}
\end{equation}
Note that $\muc=\muct$ for $p\leq 2$.

While most of these results can be evaluated also for $\alpha=0$, the case $p\to\infty$ is special. Here, Eq.~\eqref{eq:binomial-approximation-self} gives $x=\mu/(1-2\mu)$ if we again approximate $(1-2x)^L\approx \e^{-2xL}$, hence the error threshold is $\muc=1/6$, independent of $\beta$ and $L$.

\section{\label{app:cross-specific}Cross-specific replication}
\subsection{\label{app:cross-derivation}Derivation of an equation for $a$}

To obtain an equation for the parameter $a$, we compute the second moment $\sum_k k^2 \dot x_k$ of the reduced rate equations{, Eq.~\eqref{eq:rate-equations-reduced},} under the assumption that $x_k = x_k^\text{c}\equiv\frac{1}{2}\binom{L}{k}[a^k (1-a)^{L-k} + a^{L-k} (1-a)^k]$ is a sum of two complementary binomials, because in this case the first moment vanishes by construction. The four terms $T_1$-$T_4$ are defined and evaluated analogously to Eq.~\eqref{eq:T1T4}, and after some algebra we find:
\begin{align}
T_1 =&\ \frac{1}{2}\alpha L\left[L-2(L-1)(a(1-a)(1-2\mu)^2 + \mu(1-\mu))\right] \\
T_2 =&\ \frac{1}{2}\beta \Big[L(L-2(L-1)(a+\mu(1-2a))(1-a-\mu(1-2a)))   C(1-2a(1-a)) \nonumber \\  &\ + 2 a(1-a)(1-2a)^2(1-2\mu)^2
(L-1)C'(1-2a(1-a)) \nonumber \\ &\ +2(a(1-a)(1-2a)(1-2\mu))^2 C''(1-2a(1-a))\Big] \\
T_3 =& -\frac{1}{2}\alpha L[L-2(L-1)a(1-a)]\\
T_4 =& -\frac{1}{2}\beta L[L-2(L-1)a(1-a)]C(1-2a(1-a)).
\end{align}
Here, we have defined $C(x)=\frac{1}{2}\sum_k\binom{L}{k}\big[f_\text{c}(k)+f_\text{c}(L-k)\big]x^{L-k}(1-x)^k$. Adding up $T_1+T_2+T_3+T_4=0$ gives the condition
\begin{multline}\label{eq:binomial-approximation-cross}
(1-2a)^2\Bigg\lbrace \mu(1-\mu)L(L-1) \big[\alpha+\beta C(1-2a(1-a))\big] \\
-\beta a(1-a) (1-2\mu)^2\big[(L-1)C'(1-2a(1-a))\\+a(1-a)C''(1-2a(1-a))\big]\Bigg\rbrace=0.
\end{multline}

Most importantly, Eq.~\eqref{eq:binomial-approximation-cross} reads for $p=1$ in the specificity function $f_\text{c}(d)=(d/L)^p$:
\begin{equation}
-(1-2a)^2 L(L-1)\mu(1-\mu)(\alpha+\beta/2)=0,
\end{equation}
which has only the solution $a=1/2$.

\subsection{\label{app:cross-bifurcation}Bifurcation diagram}

Solving Eq.~\eqref{eq:binomial-approximation-cross} for $\mu$ gives the bifurcation diagram:
\begin{equation}\label{eq:bif-diag-cross}
\mu(a)=\tfrac{1}{2}\left[1\pm\left(1+\tfrac{4\beta a(1-a)\left[(L-1)C'(1-2a(1-a))+a(1-a)C''(1-2a(1-a))\right]}{L(L-1)\left[\alpha+\beta C(1-2a(1-a))\right]}\right)^{-1/2}\right],
\end{equation}
where we take the negative sign and the positive root to obtain values $\mu$ near zero (values near unity imply complementary replication and give equivalent results for cross-specific replication).

We readily find that $\mu''(1/2) > 0$ if $\beta > 0$, which implies that the pitchfork bifurcation described through Eq.~\eqref{eq:bif-diag-cross} is always subcritical.

\subsection{\label{app:cross-solutions}Solutions of Eq.~\eqref{eq:binomial-approximation-cross}}

The auxiliary function $C(x)$ can be evaluated for small integer $p$ as in Eq.~\eqref{eq:instances-S}. There is no solution to Eq.~\eqref{eq:binomial-approximation-cross} except $a=1/2$ for $p=1$, and the expression for $p=2$ is quite lengthy. For $p=\infty$ and $\alpha > 0$, we get
\begin{equation}
x=a(1-a)=\mu-\frac{1}{2L}\mathcal{W}\left(\frac{4\alpha\mu L}{\beta}\e^{2\mu L}\right),
\end{equation}
which is exactly the result for the self-specific case if we replace $\beta\to\beta/2$. Accordingly, we get $\muc=\mathcal{W}[\beta/(2\e\alpha)]/(2L)$ for the error threshold.

Observing that $C'(1/2)=0$, the critical mutation rate $\muct=\mu(1/2)$ is given by
\begin{equation}
\muct = \frac{1}{2}\left[1-\left(1+\frac{\beta C''(1/2)}{4 L(L-1)(\alpha+\beta C(1/2))}\right)^{-1/2}\right],
\end{equation}
which reads explicitly
\begin{equation}
\muct = \begin{cases}
0,&p=1\\
\frac{1}{2}\left[1-\left(1+\frac{2\beta}{4\alpha L^2+\beta L(L+1)}\right)^{-1/2}\right],&p=2\\
\frac{1}{2}\left[1-\left(1+\frac{\beta}{\alpha 2^L+\beta}\right)^{-1/2}\right],&p=\infty.
\end{cases}
\end{equation}

We can simply take $\alpha= 0$ in most of the above expressions to investigate the case of zero non-enzymatic replication rate. In the limits $p\to\infty$ and $L\to\infty$, we find that the bifurcation diagram $\mu(a)$ converges towards 
\begin{equation}
\mu(a)\to a(1-a),
\end{equation}
except for a region near $a=1/2${, because $\mu(1/2)$ has to coincide with the exact value  $\muct=(2-\sqrt 2)/4$.} Because this region becomes infinitesimally small as $L\to\infty$, we conclude that $\muc\to 1/4$ in this limit.


\end{document}